\documentclass[11pt, a4paper]{article}
\usepackage{moriond,epsfig}

\def\be{\begin{equation}}
\def\ee{\end{equation}}
\def\bea{\begin{eqnarray}}
\def\eea{\end{eqnarray}}

\begin{document}
\vspace*{4cm}
\title{S~Ori~70: still a strong cluster planet candidate}

\author{ E.L. Mart\'\i n (1)}
\address{(1) Instituto de Astrof\'\i sica de Canarias\\
Avda. V\'\i a L\'actea, E-38200 La Laguna, Spain}
%

\maketitle\abstracts{ In this paper I show that the coolest   
$\sigma$Orionis cluster planet S~Ori~70 is still a strong candidate member 
despite recent claims by Burgasser et al. that it could be a brown dwarf interloper. 
The main point of my argument is that 
the colors of S~Ori~70 are significantly different to those of field dwarfs. 
This object has in fact the reddest $H-K$ color of all known T dwarfs, a clear indication 
of low gravity according to all published models.  
In a $J-H$ versus $H-K$ diagram, S~Ori~70 lies in the 
region where models of ultracool dwarfs predict that low gravity objects 
should be located. I conclude that S~Ori~70 is still a strong candidate member 
of the  $\sigma$Orionis open cluster. 
I briefly discuss additional observational tests that 
can be carried out with existing facilities to verify the 
$\sigma$Orionis membership of this cluster planet candidate.}
\noindent
{\small¥{\it Keywords}: stars: very low mass stars and brown dwarfs.}

\section{Introduction}

The term brown dwarf (BD) refers to objects with masses below the limit for 
stable hydrogen fusion in stellar interiors. For solar composition, 
this limit was calculated to be 0.08~M$_\odot$ by Kumar (1963) and 
Hayashi \& Nakano (1963). 
Modern calculations have changed the value of this boundary only slightly.  
For example, Baraffe et al. (1998) give 0.072~M$_\odot$ for solar metallicity. 

There is no consensus in the community about the minimum mass of 
brown dwarfs (Boss et al. 2003). Some argue for the deuterium burning limit at 
13 Jupiter masses (for solar composition). Some prefer a limit at around 10 Jupiter 
masses, where there appears to be a sharp rise in the number of extrasolar planets 
detected by high-precision radial velocity surveys around main-sequence stars. 
Finally, it has also been proposed to use a limit at 5 Jupiter masses where the 
mass-radius relationship of substellar objects changes its sign. 
In this paper we continue to use the deuterium limit as the mass boundary between 
brown dwarfs and planets because we do not have any strong reason to change the 
nomenclature adopted in our previous papers. 

Another term used in this paper is that of ultracool dwarf, which refers to small 
objects with very cool effective temperatures. Since 1997, two new ultracool 
spectral classes have been adopted to extend the classical OBAFGKM system into 
cooler temperatures. The L dwarfs are characterized by weak or absent TiO bands, and 
very broad NaI and KI lines in the optical spectrum 
(Mart\'\i n et al. 1997, 1999; Kirkpatrick et al. 1999, 2000). 
The T dwarfs are characterized by methane bands in the near-infrared spectra 
(Oppenheimer et al. 1995; Burgasser et al. 2002; Geballe et al. 2002). 
Current estimates of the temperatures of ultracool dwarfs range from about 2400 K 
to 1400 K for L dwarfs, and from 1400 K to 700 K for T dwarfs (Basri et al. 2000; 
Vrba et al. 2004). 

Most of the known ultracool dwarfs have been identified in the general field by the 
wide area surveys 2MASS and SDSS. These objects consist of a mixed population of very low-mass 
stars, brown dwarfs and free-floating planets formed in different star-formation events 
during the lifetime of the Milky Way. Their individual ages, chemical compositions 
and masses are not known. Our best chance to study a population of ultracool dwarfs 
of known age, chemical composition, and uniform distance is to find them in open 
clusters where the stellar populations are well characterized.  Two of the first 
brown dwarfs identified were located in the Pleiades open cluster (PPl~15 and Teide~1, 
Stauffer et al. 1994; Basri et al. 1996; Rebolo et al. 1995). 
Now, we know several dozens of bona fide brown dwarfs in the Pleiades and in other 
open clusters. 

The $\sigma$Orionis open cluster has been a region where our group has concentrated many 
efforts to reveal the substellar population (see Zapatero Osorio et al. 2003 for 
a recent review). It offers several advantages: (1) 
It is young (3-8 Myr) and, thus, the substellar objects are relatively bright and hot 
(B\'ejar et al. 2001), 
but not so young that the theoretical models cannot be reliably used to obtain 
masses (Baraffe et al. 2001). (2) It has very little extinction (Lee 1968), probably because 
the parental cloud has been blown away by the O-type star at the center of the cluster. 
(3) It is relatively nearby (distance 350 pc). (4) It is moderately rich and dense. 
B\'ejar et al. (2004, in preparation) estimate a peak central density of 0.2 members 
per square arcminute.  

So far the coolest and faintest candidate member that we have found in the 
$\sigma$Orionis cluster is the T dwarf candidate S~Ori~70 (Zapatero Osorio et al. 2002). 
It was found in a pencil-beam 
deep mini-survey of only 55 square arcminutes with a sensitivity of 21 magnitude in the 
$J$ and $H$-bands carried out with the 4.2-meter William Herschel Telescope in the 
Observatorio del Roque de los Muchachos. Follow-up near-infrared photometry and 
low-resolution spectroscopy were obtained with NIRC at the 10-meter Keck I telescope. 
The photometry is summarized in Table 1, together with the data for field T dwarfs of 
similar spectral type obtained from the literature. A mid-resolution spectrum 
in the K-band obtained with NIRSPEC on Keck II was presented in 
Mart\'\i n \& Zapatero Osorio (2003). We have claimed that both the NIRC and the NIRSPEC 
spectra are best fitted with synthetic spectra with low gravity (log~g=3.5), which is 
consistent with membership in the $\sigma$Orionis open cluster. We have estimated 
a mass of 3 Jupiter masses for an age of 3 Myr, making S~Ori~70 the least massive object 
observed directly outside the solar system.

\begin{table}[t]
\caption{Comparison of photometric data of field T4-T8 dwarfs with S~Ori~70}
\vspace{0.4cm}
\begin{center}
\begin{tabular}{|c|c|c|c|l|l|}
\hline
& & & & & \\
Name & SpT & $J$ & $J-H$ & $H-K$ & Ref. \\ 
\hline
2MASS 2254+3123 & T4 & 15.01$\pm$0.03 & 0.06$\pm$0.04 & -0.08$\pm$0.04 & K04 \\
SDSS  0000+2554 & T4.5 & 14.73$\pm$0.05 & -0.01$\pm$0.06 & -0.08$\pm$0.04 & K04 \\
SDSS  0207+0000 & T4.5 & 16.63$\pm$0.05 & -0.03$\pm$0.07 &  0.04$\pm$0.07 & L02 \\
SDSS  0926+5847 & T4.5 & 15.47$\pm$0.03 &  0.05$\pm$0.04 & -0.08$\pm$0.04 & L02 \\
2MASS 0559-1404 & T4.5 & 13.57$\pm$0.03 & -0.07$\pm$0.04 & -0.09$\pm$0.04 & L02 \\
SDSS  0742+2055 & T5   & 15.60$\pm$0.03 & -0.35$\pm$0.04 & -0.11$\pm$0.04 & K04 \\
SDSS  0830+0128 & T5.5 & 15.99$\pm$0.03 & -0.18$\pm$0.04 & -0.21$\pm$0.04 & K04 \\
SDSS  0741+2351 & T5.5 & 15.87$\pm$0.03 & -0.25$\pm$0.06 &  0.00$\pm$0.07 & K04 \\
2MASS 2339+1352 & T5.5 & 15.81$\pm$0.03 & -0.19$\pm$0.04 & -0.17$\pm$0.04 & K04 \\
{\bf SOri70}          & T5.5   & 20.28$\pm$0.10 & -0.14$\pm$0.15 & 0.64$\pm$0.25 & MZO02 \\
Gl 229B         & T6   & 14.33$\pm$0.05 & -0.02$\pm$0.07 & -0.07$\pm$0.07 & L99 \\
SDSS  1231+0847 & T6   & 15.14$\pm$0.03 & -0.26$\pm$0.04 & -0.06$\pm$0.04 & K04 \\
SDSS  2124+0100 & T6   & 15.88$\pm$0.03 & -0.24$\pm$0.04 &  0.05$\pm$0.04 & K04 \\
2MASS 0937+2931 & T6   & 14.29$\pm$0.03 & -0.38$\pm$0.04 & -0.72$\pm$0.04 & K04 \\
2MASS 1225-2739 & T6   & 14.88$\pm$0.03 & -0.29$\pm$0.04 & -0.11$\pm$0.04 & L02 \\
SDSS  1110+0116 & T6   & 16.12$\pm$0.05 & -0.10$\pm$0.07 &  0.17$\pm$0.07 & L02 \\
2MASS 1047+2124 & T6.5 & 15.46$\pm$0.03 & -0.37$\pm$0.04 & -0.37$\pm$0.04 & K04 \\
SDSS  1758+4633 & T7   & 15.86$\pm$0.03 & -0.34$\pm$0.04 &  0.08$\pm$0.04 & K04 \\
2MASS 1217-0311 & T8   & 15.56$\pm$0.03 & -0.42$\pm$0.04 &  0.06$\pm$0.04 & L02 \\
Gl 570D         & T8   & 15.33$\pm$0.05 &  0.05$\pm$0.10 &  0.01$\pm$0.19 & B00 \\
2MASS 0727+1710 & T8   & 15.19$\pm$0.03 & -0.48$\pm$0.04 & -0.02$\pm$0.04 & K04 \\
 & & & & & \\ 
\hline
\end{tabular}
\end{center}
\end{table}

\section{A critique of Burgasser et al.'s paper}

Burgasser et al. (2004) have carried out an independent analysis of our Keck 
NIRC and NIRSPEC data of S~Ori~70. We sent them our raw and reduced spectra. 
Burgasser et al. reprocessed our spectra using their own software 
and found the same results as us for S~Ori~70 but not for the comparison object 
2MASS J055919-1404. Apparently, our NIRSPEC data reduction for the 2MASS comparison object was 
incorrect, and it turned out to be a field star rather than a T dwarf. 
 
Burgasser et al. compared the spectra of S~Ori~70 
with spectra of field T brown dwarfs (bdTs) obtained by them, 
and claimed to find a good match between our object and old brown dwarfs. 
In their Figure 2, they showed a comparison of their reduction of 
our NIRSPEC spectrum of S~Ori~70 and their 
NIRSPEC spectrum of the field bdT7 2MASS1553+1532. In their Figure 4, 
they showed a comparison of our NIRC spectrum of S~Ori~70 with their 
NIRC spectrum of the field bdT6.5 2MASS1047+2124. From these comparisons they claimed 
that S~Ori~70 is a field brown dwarf that coincidentally lies in the line of sight of 
the cluster. If the interpretation of Burgasser et al. is correct, 
S~Ori~70 should be  at a distance of only 75 to 100 pc, instead of 350 pc. 

We note that contrary to the claims of Burgasser et al., there are significant differences 
between S~Ori~70 and field T dwarfs. The KI doublet at 1.25 microns is stronger in S~Ori~70  
than in  2MASS1553+1532, the pseudocontinuum bump between 1.57 and 1.62 microns 
is narrower in SOri70 than in 2MASS1047+2124, and the pseudocontinuum bump between 
2.0 and 2.2 microns is higher and redder in SOri70 than in 2MASS1047+2124. All these three 
features are consistent with the model predictions for gravity effects as also 
noted by Lucas et al. (2001) in their analysis of infrared spectra of L dwarfs 
in the Trapezium cluster and by Knapp et al. (2004) in their analysis of field T dwarfs. 
Burgasser et al. neglected to comment on the NIRC spectra mis-matches between SOri70 and 
field T dwarfs, and dismissed the detection of the KI doublet because the NIRSPEC spectrum 
is too noisy.  

Knapp et al. (2004) have proposed that the spread of $H-K$ colors observed in field 
T dwarfs could be due to differences in gravities, expected for a sample of 
brown dwarfs with a wide range of ages and masses. The models indicate that 
objects with red H-K colors have low gravities due to weaker pressure-induced H$_{\rm 2}$ 
absorption in the K-band. This effect was first predicted by Saumon et al. (1996), 
and can also be noticed in the models published by Allard et al. (2001). It seems to be 
a robust prediction of all models. 
In Figure 1, I display a color-color diagram comparing  
the position of S~Ori~70 with that of field T4-T8 dwarfs. 
The S~Ori~70 photometry was calibrated in the same system as that of Leggett et al. (2002). 
The plot includes six bdTs  with photometry from Leggett et al. (2002). 
The MKO system used by Knapp et al. (2004) is very close to the Leggett et al. system, 
and hence I have made no corrections. 
I also show the models used by Knapp et al. Full explanation of these cloudless 
models can be found in Marley et al. (2002).   
All the field bdT objects are located in the region where 
the models give gravities in the range log~g=4.0-5.5, consistent with their presumed 
old ages. S~Ori~70, on the other hand, is located outside the locus of the field 
bdTs, even taking into account the photometric error bars. 
Its red $H-K$ color indicates lower gravity than the field objects according 
to the models. In fact, Mart\'\i n \& Zapatero Osorio (2003) derived log~g=3.5 for S~Ori~70, 
which implies a mass of 3 Jupiters. The red $H-K$ color cannot be due to insterstellar 
reddening because extinction is very low in this line of sight, and because reddening 
would also affect the $J-H$ color.

\begin{figure}
\vskip 2.5cm
\psfig{figure=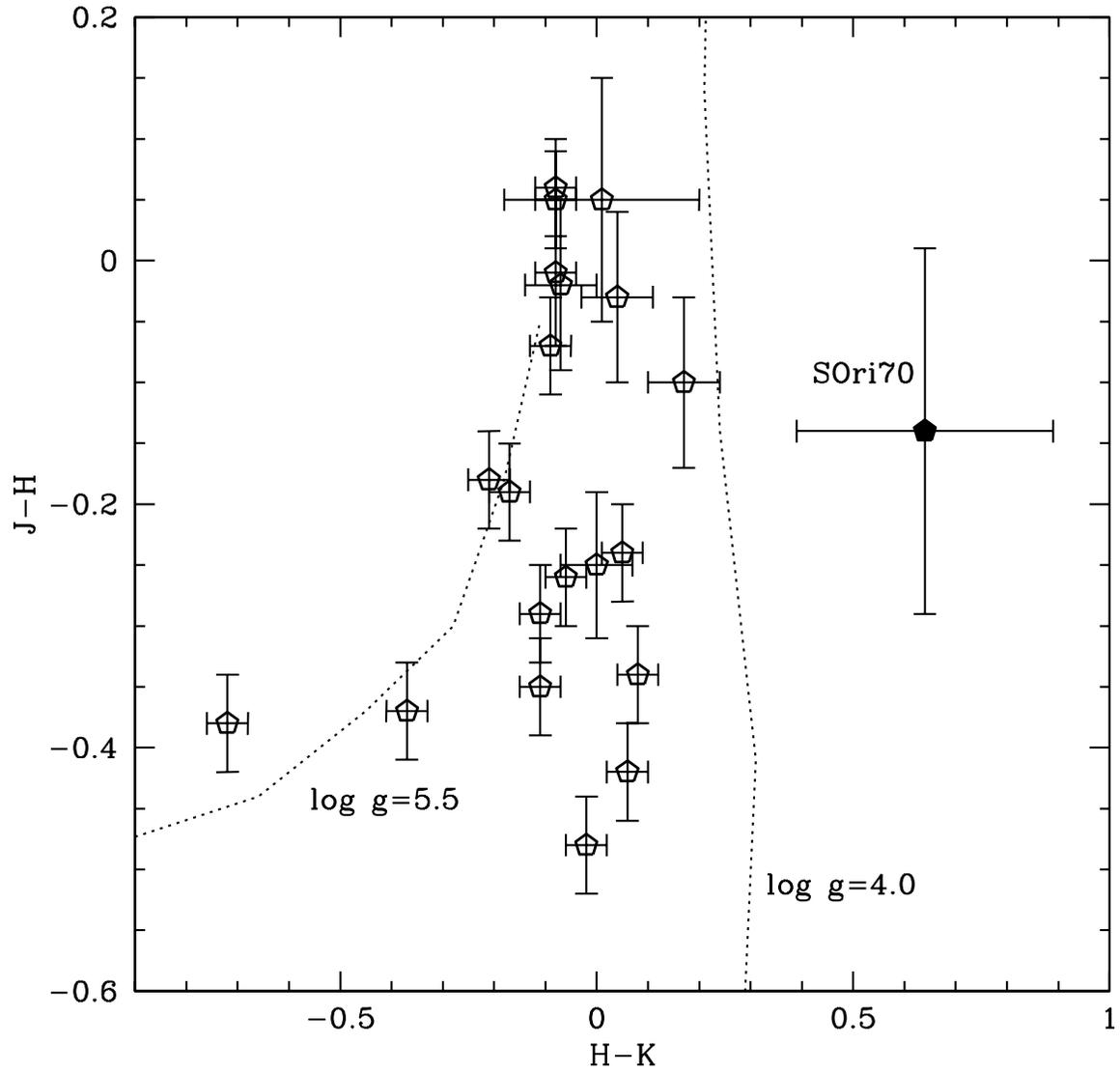,height=16cm}
\caption{Comparison of S~Ori~70 with field bdT4-T8 objects. Theoretical 
models from Marley et al. (2002) are also shown with dotted lines for two different gravities. 
Note that S~Ori~70 is located in a region outside the locus of field T 
dwarfs, consistent with having lower surface gravity. All the known 
field T dwarfs are bracketed by the models.  
\label{fig:radish}}
\end{figure}


\section{Where do we go from here?}

The differences in colors and spectral energy distribution between SOri70 and field 
bdTs strongly suggest that S~Ori~70 has lower gravity, consistent with young age 
and low mass. We conclude that, despite Burgasser's claims, S~Ori~70 is still 
a strong cluster planet candidate. Nevertheless, more data is necessary to improve our 
understanding of this object and to confirm its cluster membership in $\sigma$Orionis. 
In this section, I discuss future observations 
that will provide crucial information about S~Ori~70. 

\begin{itemize}

\item {\bf The proper motion:}  
Zapatero Osorio et al. (2002) compared images with a 3 year baseline of the field around S~Ori~70. 
They placed an upper limit of 0.1 arcsec per year on the proper motion of S~Ori~70. 
If S~Ori~70 is a bdT at 75-100 pc, the proper motion should be about 0.05 arcsec per year. 
The last images of S~Ori~70 were obtained at Keck I on December 2001. We plan to obtain new images 
in winter 2004 in order to improve the limit on the proper motion of S~Ori~70 by a factor of two. 
Unfortunately, proposals to obtain accurate astrometry of S~Ori~70 
with HST in Cycles 11, 12 and 13 have not been successful.  

\item {\bf The KI near-infrared lines:}
The detection of the KI doublet at 1.25 microns reported by  Mart\'\i n \& Zapatero Osorio (2003) 
needs to be confirmed with higher signal to noise ratio. A proposal to reobserve S~Ori~70 with 
Keck/NIRSPEC was rejected by the NASA TAC given ``the extensive study of S~Ori~70 by Burgasser and 
his colleagues''. In this paper we have shown that the Burgasser et al. study did not provide 
the final word on the nature of this object, and that 
S~Ori~70 clearly deserves further investigation. 

\item {\bf The spectral energy distribution (SED):}  
The SED coverage of S~Ori~70 is still incomplete and the error bars in the photometry should 
be reduced. 
A proposal to obtain the $JH$ spectrum with Keck/NIRC, 
and to improve the $HK$ spectrum was rejected by the NASA TAC. A proposal to obtain mid-infrared 
photometry with SIRTF/IRAC was also turned down. We plan to insist on applying for telescope time 
to observe S~Ori~70 so that we can obtain an excellent SED of this benchmark cluster planet, 
which can be 
a reference for future studies of planetary-mass objects in star-forming regions and young 
open clusters. In parallel with more detailed observations of  S~Ori~70, there should 
also be theoretical efforts to improve the synthetic spectra of T dwarfs. The best fit 
of the J-band  NIRSPEC spectrum found by Mart\'\i n \& Zapatero Osorio (2003) has a 
T$_{\rm eff}$=1,100 K, significantly hotter than the best fit of the $HK$ NIRC spectrum 
for T$_{\rm eff}$=800 K obtained by Zapatero Osorio et al. (2002). Burgasser et al. (2004) 
correctly pointed out this inconsistency in the comparison of observations with 
theory and also showed that this fitting 
technique provides gravities that are too low for bdT7-bdT8 field objects. However, 
the gravities obtained from the models are reasonable for bdT6 field objects, which is closer 
to the spectral type of S~Ori~70.  

\end{itemize}

Last but not least, I would like to note that 
our efforts to find more cluster planets similar to S~Ori~70 have met with poor weather in 
several observatories (Calar Alto, Canaries, Hawaii and Paranal). 
We do not have a reliable database to search for these objects yet. 
Nevertheless, we have found a few candidates that await spectroscopic observations. 
I showed one of them at the Moriond meeting, a possible planetary mass companion to 
an M dwarf member in the $\sigma$Orionis open cluster.  
Follow-up observations of these objects with several telescopes are planned. Stay tuned!

\section*{Acknowledgments}
Partial funding for attending this conference was provided by a grant 
from the RTN network ``The Formation and Evolution of Young Stellar Clusters'' 
led by Mark McCaughrean. Adam Burgasser, Jose Antonio Caballero,  David Barrado y Navascu\'es 
and Maria Rosa Zapatero Osorio 
provided useful comments that helped to improve the manuscript. Mark Marley sent a computer 
readable version of the models used in Figure 1.

\end{document}